\documentclass[sigconf]{acmart}
\usepackage{color}
\usepackage{soul}
\usepackage{microtype}
\usepackage[T1]{fontenc}
\usepackage{array}
\usepackage{subcaption}
\AtBeginDocument{%
  }

\copyrightyear{2026}
\acmYear{2026}
\setcopyright{cc}
\setcctype{by}
\acmConference[RecSys '26]{20th ACM Conference on Recommender Systems}{September 27-October 02, 2026}{Minneapolis, MN, USA}
\acmBooktitle{20th ACM Conference on Recommender Systems (RecSys '26), September 27-October 02, 2026, Minneapolis, MN, USA}
\acmDOI{10.1145/3773078.3831900}
\acmISBN{979-8-4007-2284-4/2026/09}

\begin{document}

% \title{SiDec: Decode Semantic IDs for Recommendation Systems}
\title[Tokens are All You Need: Dual-purpose Semantic IDs]{Tokens are All You Need: Dual-purpose Semantic IDs for Achieving LLM-Level I/O Efficiency in Recommendation Systems}


% --- Authors ---
\author{Baolei Li}
\authornote{Both authors contributed equally to this research.}
\affiliation{%
   \institution{YouTube}
   \country{}
}
\email{baoleili@google.com}

\author{Yiping Yuan}
\authornotemark[1]
\affiliation{%
   \institution{YouTube}
   \country{}
}
\email{yipingyuan@google.com}

\author{Yilin Zheng}
\affiliation{%
   \institution{YouTube}
   \country{}
}
\email{yilinzheng@google.com}

\author{Likang Yin}
\affiliation{%
   \institution{YouTube}
   \country{}
}
\email{lkyin@google.com}

\author{Ling Liu}
\affiliation{%
   \institution{YouTube}
   \country{}
}
\email{liuling@google.com}

\author{Fabio Soldo}
\affiliation{%
   \institution{YouTube}
   \country{}
}
\email{fsoldo@google.com}

\author{Romer Rosales}
\affiliation{%
   \institution{YouTube}
   \country{}
}
\email{romerrosales@google.com}

\author{Xinyang Yi}
\affiliation{%
   \institution{Google Deepmind}
   \country{}
}
\email{xinyang@google.com}

\author{Lichan Hong}
\affiliation{%
   \institution{Google Deepmind}
   \country{}
}
\email{lichan@google.com}

\renewcommand{\shortauthors}{B. Li, Y. Yuan et al.}

\begin{abstract} 
Large-scale recommendation systems face "Memory Wall" bottlenecks due to massive, dense embedding tables.
While generative retrieval uses discrete tokens for IDs, high-dimensional context still relies on inefficient dense formats. Inspired by computer vision data compression, we propose Dual-purpose Semantic IDs to achieve LLM-level I/O efficiency.
Our methodology uses hierarchical quantization to condense continuous embeddings into discrete Semantic IDs performing two concurrent roles: (1) Collaborative Identity: modeling user-item interactions via learnable embedding table; and (2) Content Reconstruction: using a lightweight Semantic Decoder for on-the-fly embedding approximation. This approach replaces massive vector storage with on-demand reconstruction, reducing system overhead and data footprints. We demonstrate the efficacy of
our framework through offline evaluations and successful online deployment in production-scale ranking and retrieval
systems at a major video sharing platform, showing that discrete
tokens are indeed all you need for highly efficient, content-rich
recommendation.
\end{abstract}





\maketitle

\section{Introduction}
In recent years, deep recommendation systems have sought to mirror the scaling laws observed in Large Language Models (LLMs).
However, while LLMs scale elegantly due to their compute-bound nature and unified discrete token space, large recommendation systems remain fundamentally restricted by the "Memory Wall." Modern architectures rely heavily on vast, dense floating-point embedding tables to represent users, items, and high-dimensional continuous features. Ingesting these massive numerical vectors during training and inference creates immense I/O and memory bandwidth bottlenecks, fundamentally capping the throughput and serving efficiency of large-scale recommendation models. This is particularly challenging when the recommendation system evolves towards a more sequential formation and handling longer user activity sequences with sequence lengths at scale of $10^4$ or higher \cite{10.1145/3705328.3748065}.

To circumvent the limitations of traditional dot-product-based dual encoders, a paradigm shift toward generative retrieval has emerged \cite{rajput2023recommender, he2026plum}. In these works, semantic tokens were aligned with text tokens to utilize the prediction power of pre-trained LLMs. However, the focus has predominantly been on using these semantic tokens strictly as replacements for categorical item or video IDs. The handling of high-dimensional, continuous numerical inputs—such as complex contextual signals, historical engagement densities, and pre-trained content embeddings—remains tethered to inefficient, dense continuous representations.

To break free from this, we look to the evolution of computer vision. Historically, processing continuous, high-dimensional spatial data (pixels) was a bottleneck for generative models. This paradigm was elevated by the introduction of VQ-VAE \cite{oord2017neural} and subsequently VQGAN \cite{esser2021taming}.
These works revolutionized the field by proving that continuous spatial data could be effectively compressed into a sequence of discrete tokens without sacrificing intrinsic semantic meaning. By quantizing the continuous space into a learned codebook, VQGAN allowed standard Transformer architectures to process images with the same efficiency as text. The implication is profound: high-dimensional, continuous distributions do not need to be processed in their native floating-point format to retain their predictive power.

The contributions of this work are twofold: 
\begin{itemize}
    \item We introduce a novel dual-purpose framework that addresses the "Memory Wall" and I/O bottlenecks through extreme embedding compression. By integrating standard Semantic ID embedding learning with on-the-fly Semantic ID Decoding (SiDec), our approach achieves an optimal balance between item-specific memorization (via discrete tokens) and content-aware generalization (via reconstructed continuous semantics).
    \item We provide extensive empirical evidence of this framework's effectiveness within a production-scale recommendation system. Through rigorous offline benchmarking and live online deployments, we demonstrate that dual-purpose Semantic IDs drastically reduce data footprints and system overhead while delivering substantial improvements in recommendation quality.
\end{itemize}

Specifically, we propose to bring the above information compression angle to leverage Semantic ID in recommendation systems, where the Semantic IDs are usually built from high-dimensional item embedding understanding from powerful content understanding models, such as output from multi-modality LLM. By decoding the semantic ID with its original codebook injected within recommendation models, we can reconstruct the deep content understanding with minimum system overhead for storing and transferring the large dimension embedding features. Furthermore, we propose to use a lightweight custom learned decoder to better align the content understanding in task specific recommendation models. This new Semantic ID Decoding (SiDec) approach complements the conventional approach of sparse embedding learning, since the reconstructed content understanding contains rich hierarchical semantic meaning with full coverage of the item corpus.

\section{Related work}

Our proposed methodology intersects several domains, including generative retrieval, discrete vector quantization, and compressed sequence modeling.

\subsection{Semantic ID and Generative Retrieval}
Semantic ID for recommendation systems has been widely explored since the introduction of Semantic ID \cite{rajput2023recommender, singh2024better}. Replacing meaningless atomic item IDs with sequences of discrete, semantically rich tokens, strikes a good balance between memorization and generalization and performs better for fresh and long-tail items (the cold-start problem). It enabled the rise of Generative Recommendation paradigm \cite{rajput2023recommender, he2026plum, deng2025onerec, zhai2024actions}. However, prior research primarily employs SIDs solely as discrete identification tokens, missing the potential to utilize them as efficient vectors for raw content signals.

\subsection{Discrete Vector Quantization in Deep Learning}
Discretizing continuous, high-dimensional spaces to enhance model efficiency has strong roots in computer vision and generative modeling. The Vector Quantized Variational Autoencoder (VQ-VAE) \cite{oord2017neural} and subsequent VQGAN \cite{esser2021taming} demonstrated that dense, spatial image embeddings could be effectively compressed into codebook-indexed discrete sequences without sacrificing fine-grained semantic details. 
These spatial quantization principles are highly applicable to modern recommender architectures, where various methods have been proposed \cite{kasalicky2025future, kusupati2022matryoshka} to process dense embedding features. Our framework follows the quantization philosophy, treating high-dimensional, continuous item embeddings as visual codebooks that can be discrete-tokenized, transmitted efficiently over the network, and reconstructed only when necessary inside the execution graph.

\subsection{Semantic IDs as Content Feature Compression}
Recent literature has rapidly expanded upon the utility of Semantic IDs, exploring techniques to leverage them for feature compression and on-the-fly reconstruction. Two contemporary paradigms are highly relevant to our work:

\begin{itemize}    
    \item \textbf{Embedding-Free Sequence Learning (SIDE):} 
    The SIDE framework \cite{ramasamy2025side} applies a novel vector quantization technique to compress multiple item-level content embeddings into ternary codewords. These codewords are packed via n-gram hashing and directly parsed into feature representations inside the model, eliminating the memory overhead of traditional embedding lookup tables. While sharing the goal of reducing table parameters, \textit{SiDec} differs from SIDE in two key aspects: First, instead of relying on custom ternary or scalar quantization layers, we utilize standard Residual Quantization (RQ-VAE) \cite{lee2022autoregressive}, which is natively adopted within the Semantic ID and Generative Retrieval community. Second, rather than projecting raw codewords from scratch, \textit{SiDec} directly leverages the latent space of the original quantization codebook to perform stable, high-fidelity embedding reconstructions.
    
    \item \textbf{Hierarchical Target Routing (HiSAC):} 
    To compress ultra-long sequential user histories, HiSAC \cite{yuan2026hisac} utilizes multi-level Semantic ID codebooks to represent personalized user interest-agents. It introduces a Soft-Routing Attention mechanism that directly decodes relationships between candidates and these compressed interest-agents, minimizing quantization loss. Our approach diverges from HiSAC along two fundamental architectural directions: First, \textit{SiDec} is designed to reconstruct content embeddings as a general-purpose, application-agnostic input feature. We do not restrict how this reconstructed feature is consumed by downstream layers (making it equally suitable for both ranking and retrieval models). Second, rather than relying on a static, pre-trained codebook decoder, \textit{SiDec} implements a trainable decoder explicitly designed to dynamically align reconstructed content semantics with target tasks during joint end-to-end training.
\end{itemize}

\section{Methodology}
The core of our approach is the transformation of high-dimensional continuous content embeddings into a compact, discrete token space that serves a dual purpose: acting as a unique identifier for collaborative filtering and as a compressed representation for content-based understanding. Below we will introduce how the Semantic ID was generated and two complementary approaches that are used in our recommendation system.

\begin{figure*}[ht]
  \centering
  \includegraphics[width=0.76\linewidth]{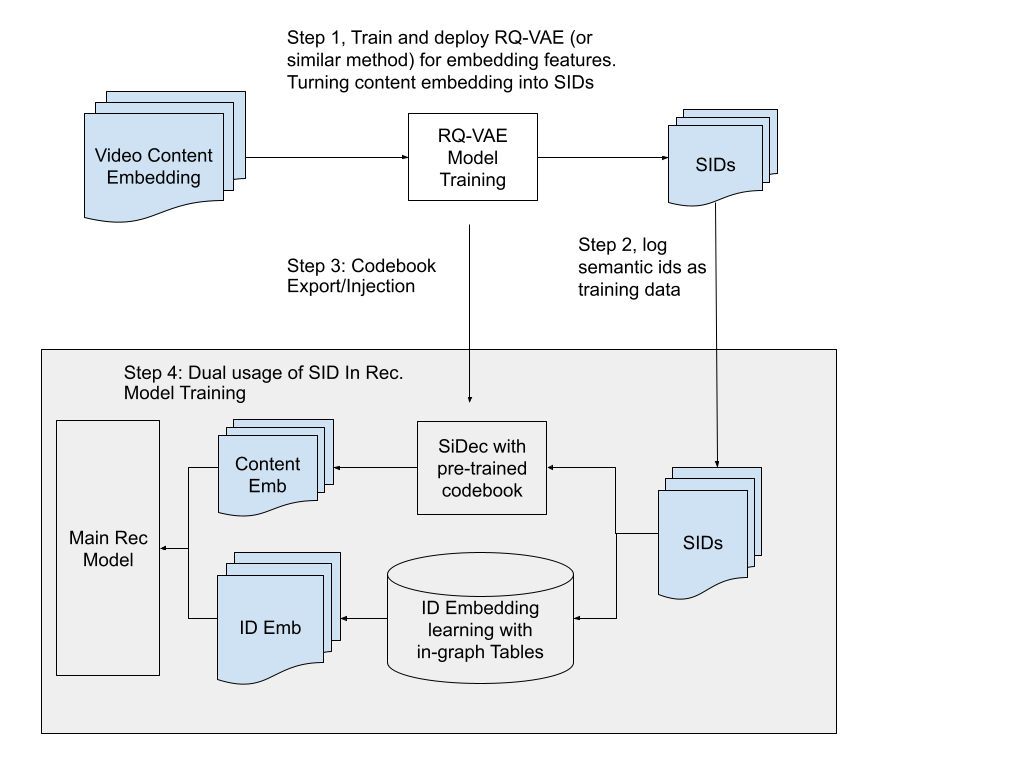}
  \caption{Semantic ID as an expressway to deliver semantic embedding
}
  \Description{Semantic ID as an expressway to deliver semantic embedding
}
\end{figure*}

\subsection{Semantic ID Generation via Quantization}

Semantic IDs are discrete tokenized codewords derived from high dimensional content features. Formally, let $\mathcal{I}$ be the set of items in our corpus. For each item $i \in \mathcal{I}$, we assume the existence of a high-dimensional content embedding $\mathbf{e}_i \in \mathbb{R}^d$, typically generated from a pre-trained content model (e.g., a video-language model or a multimodal transformer). 

To achieve I/O efficiency, we compress $\mathbf{e}_i$ into a sequence of $K$ discrete tokens using a hierarchical quantization framework, such as Residual Quantization (RQ-VAE) \cite{lee2022autoregressive, rajput2023recommender, he2026plum}. The Semantic ID for item $i$ is defined as:
\begin{equation} \label{eq:sid}
    S_i = [t_{i,1}, t_{i,2}, \dots, t_{i,K}]
\end{equation}
where each $t_{i,k} \in \{1, \dots, V\}$ is an index from a shared codebook $\mathcal{C}$ of size $V$. This discrete representation reduces the storage requirement from $d \times 32$ bits (for floating-point vectors) to $K \times \log_2(V)$ bits, typically achieving a compression ratio of 50--100$\times$.

\subsection{Dual-Purpose Semantic ID Framework}
Unlike traditional systems that treat IDs and content features as separate entities, our framework utilizes $S_i$ for two simultaneous functions within the recommendation model:

\begin{enumerate}
    \item \textbf{In-Graph Collaborative Identity Embedding Learning:} The sequence $S_i$ is treated as a set of categorical features. The model learns embeddings for each token $t_{i,k}$, allowing it to capture user-item interaction patterns. By sharing prefixes among semantically similar items, the model naturally generalizes across the item cold-start boundary.
    \item \textbf{Semantic Decoding (SiDec):} To capture the "pure" content signal without the I/O cost of joining dense features, we introduce an operator $\phi$ to perform codebook lookup for each token and aggregate their embedding from the codebook, as well as a \textit{Semantic Decoder} $f_\theta$. They, together, reconstruct an approximation of the original content embedding:
    \begin{equation} \label{eq:sidec}
        \hat{\mathbf{e}}_i = f_\theta(\phi(S_i)) 
    \end{equation}
\end{enumerate}

\subsection{In-Graph Collaborative Identity} \label{sec:collabrative_token}
While the SiDec approach captures the content-centric signal, the recommendation model must also learn the collaborative "identity" of an item from user interactions. Traditional systems use a single unique embedding for each item ID, which fails to generalize to new or tail items. In our framework, we decompose the Semantic ID $S_i = [t_{i,1}, t_{i,2}, \dots, t_{i,K}]$ into various token-based features to balance memorization (learning specific item behavior) and generalization (sharing patterns across semantically similar items).

We consider four primary strategies for embedding $K-$token Semantic IDs within the model graph:

\subsubsection{Unigram Representation}
The simplest approach treats each token as an independent categorical feature. For an item $i$, the representation is the aggregation of individual token embeddings:
\begin{equation}
    \mathbf{x}_i^{uni} = \text{Aggregate}(\{\text{Emb}_k(t_{i,k}) \mid k=1, \dots, K\})
\end{equation}
where $\text{Emb}_k(t_{i,k})$ is the $k-$th lookup embedding indexed by the $k-$th token of the semantic ID. This approach has the lowest memory footprint but lacks the ability to capture the specific dependency between hierarchy levels. $\text{Aggregate}$ can be a summation or concatenation operator.

\subsubsection{Overlapping Bigram Representation}
To capture local transitions within the semantic space without being strictly tied to the root of the hierarchy, we employ overlapping bigrams (sliding window):
\begin{equation}
    \mathbf{x}_i^{over} = \text{Aggregate}(\{ \text{Emb}_{k, k+1}(t_{i,k}, t_{i,k+1}) \mid k=1, \dots, K-1 \})
\end{equation}
This approach provides good balance between generalization and memorization. Bigram vocabs are usually tractable. It is particularly effective at capturing "semantic neighborhoods" that may span across different sub-clusters, offering a more flexible relational structure that can be more robust to minor noise in the quantization process.

\subsubsection{Sentence Piece Model (SPM)}
In addition to the fixed representation strategy above, we can also use the more adaptive method like Sentence Piece Model (SPM) \cite{kudo2018subword, singh2024better}. SPM learns the representation based on the empirical data distribution to automatically balance how the tokens are combined.
\begin{equation}
    \mathbf{x}_i^{spm} = \text{Emb}(\{ \text{SPM}(t_{i,1}, \dots, t_{i,k}) \mid k=1, \dots, K \})
\end{equation}

\subsubsection{Nested N-gram (Hierarchical) Representation}
To explicitly model the hierarchical nature of Semantic IDs—where the first token represents a coarse cluster and subsequent tokens represent finer refinements—we use nested n-grams. Each feature is defined by the prefix of the ID:
\begin{equation}
    \mathbf{x}_i^{nest} = \text{Aggregate}(\{ \text{Emb}_{1:k}(t_{i,1}, \dots, t_{i,k}) \mid k=1, \dots, D \})
\end{equation}

This design ensures that items sharing the same semantic prefix (e.g., all "Jazz Music" videos) share the exact same top-level embeddings. The model learns collaborative signals at multiple granularities, allowing it to provide meaningful recommendations for tail items by leveraging the learned behavior of their parent semantic clusters. It provides flexibility to control the level of memorization through the n-gram vocab sizes. We can set $D\leq K$ to limit the granularity. For embedding tables with $k>2$, we can allow random hashing to limit the $N-$gram vocab size.  

\subsubsection{Trade-offs and Capacity}
In our implementation, we typically combine these representations or select the strategy based on the specific use case (Ranking vs. Retrieval). The \textit{Nested N-gram} approach and \textit{SPM} are preferred for cold-start scenarios, as they enforce a strict hierarchy. It also offers more flexibility toward memorization. However, it requires a larger embedding table to store the prefixes. Conversely, \textit{Unigram} and \textit{Overlapping Bigram} approaches offer superior I/O efficiency and a smaller memory footprint by re-using token embeddings more aggressively across different positions in the sequence.

\subsubsection{Real-world Applications}\label{nested_ngram}
To illustrate the effectiveness of this collaborative ID embedding approach, we would like to share one of our practices with the Nested N-gram strategy. By leveraging the hierarchical design, we were able to completely replace item-level Video ID embeddings in our Homepage ranking model while preserving baseline performance. In online A/B testing, this approach achieved neutral top-line metrics alongside significant gains in item freshness:
\begin{itemize}
    \item \textbf{+1.11\%} Homepage 7-Day Click-Through Rate (CTR)
    \item \textbf{+0.85\%} Homepage 1-Day CTR
    \item \textbf{+0.59\%} Homepage 3-Hour CTR
\end{itemize}
These gains demonstrate that multi-granularity semantic embeddings successfully eliminate cold-start lag, allowing the recommender to immediately surface newly published content without sacrificing overall recommendation quality.

\subsection{The Semantic Decoder Architecture}\label{sec:semantic_decoder}
The decoder $f_\theta$ is designed to be lightweight to maintain inference efficiency. We employ a Multi-Layer Perceptron (MLP) or a shallow Transformer that operates on the codebook embeddings in a latent space of the tokens $\phi(S_i)$, where $\phi$ is a static codebook-lookup and summation layer. 

During the Semantic ID generation stage, the codebook and the reconstruction is trained to minimize the mean squared error (MSE) against the original content embedding:
\begin{equation}
    \mathcal{L}_{rec} = \sum_{i \in \mathcal{I}} || \mathbf{e}_i - f_\theta(\phi(S_i)) ||^2
\end{equation}
By freezing the decoder during the recommendation model training (or fine-tuning it with a small learning rate), we ensure that $\hat{\mathbf{e}}_i$ provides a stable, content-centric signal that is independent of item popularity.

\subsubsection{I/O-Efficient Model Integration}
The primary bottleneck in large-scale ranking and retrieval is the "join" operation required to attach dense content embeddings to user history, such as user's watch history.
In our methodology, we replace this with a simple lookup of the $K$ integer tokens. In particular, we replace the storage of dense content embeddings in the training logs with their corresponding $K$-token Semantic IDs. For each item $j_l \in H$ in the user history $H = \{j_1, j_2, \dots, j_L\}$, the model performs the following:
\begin{enumerate}
    \item \textbf{Token Parsing:} Retrieve the $K$ discrete tokens $S_j$ via Eq. \eqref{eq:sid}.
    \item \textbf{Codebook lookup:} Perform a codebook lookup $\phi$ to reconstruct the semantic representation in a hidden space, where the codebook is exported from the pre-trained model for Semantic ID generation.
    \item \textbf{Semantic Decoding:} Pass the hidden space representation through the pre-trained decoder $f_\theta$ to reconstruct the approximate content embedding with Eq. \eqref{eq:sidec}. Note that the decoder $f_\theta$ can be a new trainable module, to better align the semantic understanding in application domain; or it can be the identity operation, in which case the target application directly leverages the static latent representation.
    \item \textbf{Contextual Aggregation:} The reconstructed embeddings are then fed into the model's sequence processor (e.g., a Transformer encoder or a Mean Pooling layer):
    \begin{equation}
        \mathbf{u}_{history} = \text{Attention}(\mathbf{q}_{cand}, \hat{\mathbf{e}}_{j_1}, \dots, \hat{\mathbf{e}}_{j_L}),
    \end{equation}
    where $\mathbf{q}_{cand}$ is the decoded candidate item embedding via SiDec.
\end{enumerate}
With this approach, the overall cost of adding reconstructed content embedding will be equivalent to adding few sparse features (number of codebooks, typically 3 to 8) with low-cardinality embedding tables (number of rows in each codebook, few hundreds to few thousands), which is marginal to most of the modern recommendation systems. The codebooks are also kept static for given versions of Semantic ID, which usually have a lifetime spanning from months to years. While it is possible to decouple the SID versions for SiDec and Collaborative ID approaches, we usually align their versions in practice to amortize the cost of feature extraction and logging.

\subsubsection{Ranking Models} 
For a candidate item $i$ and a user history $H = \{j_1, j_2, \dots, j_L\}$, the model input becomes a sequence of Semantic IDs. The reconstructed embeddings $\hat{\mathbf{e}}_{j}$ are computed on-the-fly within the model graph. This eliminates the need to log or join $d$-dimensional vectors for every item in the history, drastically reducing the training data footprint. For example, with a user history length of $L$ items and a content embedding dimension $d$, the model must ingest $L \times d$ floating-point values per sample. For example, when $L=200$ and $d=256$, this results in $51, 200$ floats ($200$ KB in FP32) per training example. At the scale of billions of examples, the cost of logging, storing, and joining these dense vectors becomes the primary bottleneck for training throughput, and usually prohibitively costly in production. With the on-the-fly SiDec approach, the additional cost of adding such a reconstructed embedding feature is incremental, not more than adding a normal sparse embedding feature and with orders of magnitude less sparse parameters. 

\begin{table*}[htbp]
    \caption{Summary of real world application results. \textmd{Note: Reported metrics may represent proprietary variants of the listed descriptors for each application. The experiment is performed on our production AB test system with P-Value=0.05, and the percentage change is followed by a confidence interval in the square brackets.}}
    \label{tab:results_summary}
    \vspace{-10pt} % Pulls the table up closer to the caption
    \centering
    \renewcommand{\arraystretch}{1.2}
    \setlength{\tabcolsep}{3pt} % Tightened further from 4pt to 3pt
    % Re-balanced widths: Shrunk Col 2, expanded Cols 3 and 5
    \begin{tabular}{@{} >{\raggedright\arraybackslash}p{0.25\linewidth} >{\raggedright\arraybackslash}p{0.32\linewidth} >{\raggedright\arraybackslash}p{0.32\linewidth} @{}}
        \toprule
        \textbf{Applications} & \textbf{Semantic ID Features} &  \textbf{Online Satisfied Engagement} \\
        \midrule
        
        \textbf{Watchpage Ranking} & Watch, Candidate, and Watch History Semantic IDs & $+$0.09\%:[0.06, 0.11] Sitewide \newline $+$0.80\%:[0.74, 0.86] Watchpage \\
        \midrule
        
        \textbf{Homepage Ranking} & Candidate and Watch History Semantic IDs  & $+$0.08\%:[0.04, 0.12] Sitewide \newline $+$0.28\%:[0.01, 0.56] New User \newline $+$0.22\%:[0.17, 0.28] Homepage  \\
        
        \midrule
        
        \textbf{Retrieval Model}  & Watch History Semantic IDs & $+$0.06\%:[0.06, 0.11] Sitewide \newline $+$0.17\%:[0.02, 0.33] New User \newline $+$0.13\%:[0.06, 0.2] Homepage \newline $+$0.09\%:[0.0, 0.17] Watchpage \\
        \bottomrule
    \end{tabular}
\end{table*}

\subsubsection{Retrieval Models} 
A fundamental challenge in scaling retrieval models lies in video watch representation in users' watch history. (e.g. SASRec-based architectures \cite{kang2018self}) 
Traditionally, models have relied on learnable embeddings mapped to discrete item IDs (e.g., video/channel IDs) or coarse cluster tokens (e.g., video clusters). While computationally efficient and highly effective for high-frequency popular content, this approach relies heavily on accumulated co-occurrence statistics. Consequently, it suffers from severe data sparsity and popularity bias, failing to generalize well to new users or long-tail content.

Conversely, the approach that ingests pre-trained, high dimensional dense content embeddings offers better robustness to sparsity and zero-shot generalization compared to the self-learned embeddings. However, the I/O-bound nature of streaming large, pre-trained content embeddings, compounded by severe memory bandwidth constraints, renders the direct utilization of such heavy representations prohibitively expensive in production systems processing extensive user histories.

The dual-purpose framework allows the model to utilize the hierarchical structure of Semantic IDs (e.g., prefix matching) for efficient candidate generation while simultaneously leveraging the reconstructed $\hat{\mathbf{e}}_i$ to maintain a deep understanding of content-level transitions in user behavior. Therefore, we can approximate raw semantic content features for every item in the user history without storing or transmitting uncompressed float vectors.

Specifically, the decoded semantic embeddings are concatenated with other standard user and item features to form the input to a deep autoregressive network. The core architecture is based on a Transformer network utilizing relative attention mechanisms to capture multi-scale temporal dependencies.

\section{Experiments}

In this section, we discuss how dual-purpose Semantic IDs are configured and evaluated in production for large-scale ranking and retrieval models. Since abundant research already reports on the Collaborative Identity aspect \cite{singh2024better}—which is actively developed and deployed in our production systems (see \ref{nested_ngram}) — our experiments focus primarily on the SiDec (content reconstruction) aspect of the framework. We evaluate the framework both offline to quantify I/O efficiency and online to measure real-world user impact.  

\subsection{Experimental Setup}

To ensure a rigorous and fair evaluation across different models and architectures, our experimental framework isolates data volume biases from representation quality.

\textbf{Offline Fixed-Step Training:} For our offline studies, particularly when evaluating I/O bottlenecks, we adopted a fixed-step training schedule rather than the continuous training paradigm typical of production environments. Due to the substantial disparities in training throughput (steps per second) caused by the varying I/O demands of different experimental arms, time-bounded continuous training would lead to faster models processing significantly more data and receiving more gradient updates. By enforcing a strict and identical step budget for all trainers, we guarantee that all models are exposed to the exact same compute budget. 

\textbf{Evaluation Metrics:} Offline model quality is measured using next-item prediction loss (Cross-Entropy) at convergence and candidate retrieval accuracy (Hit Rate @100) for retrieval models and Click-Through Rate (CTR) Area Under the Curve (AUC) for ranking models. System efficiency is measured via Training Speed (steps/s). Online performance is measured using \textit{Online Satisfied Engagement}, a proprietary composite metric reflecting prolonged user watch time and positive interactions (e.g., likes, saves) on the platform.

\subsection{Online Production Deployment}

We deployed the dual-purpose Semantic ID framework in both a multitask production ranking model \cite{singh2024better} and a foundational transformer retrieval model. 

For the ranking model, we utilized three key Semantic ID features: the candidate video being ranked, the video currently being watched, and the user's watch history. For the collaborative identity stream, we adopted the nested N-gram or SPM as outlined in Section~\ref{sec:collabrative_token} to provide a high-capacity collaborative signal. This allows the model to learn specific user-item biases. Concurrently, the new SiDec content reconstruction stream provides a "cold-start" friendly representation. Because the decoder $f_\theta$ is trained on a massive corpus to reconstruct raw content features, $\hat{\mathbf{e}}_j$ is robust to popularity bias and provides a stable signal even for items with zero interactions in the ranking training set.

For the retrieval model, the SiDec architecture was primarily applied to the user's watch history to safely bridge the semantic gap for ID-based embeddings without violating strict latency and resource constraints. 

We assume the production system already have the collaborative identity stream in the baseline and only measure the benefit of "dual-purpose" by introducing the SiDec. Because the discrete Semantic ID tokens are already retrieved and cached during serving for the baseline model, reconstructing the dense embeddings on-the-fly introduces negligible additional training and serving costs. (We analyze the exact I/O throughput benefits in the subsequent offline studies).

As detailed in Table~\ref{tab:results_summary}, introducing the SiDec content reconstruction stream provides highly significant top-line gains across both watchpage and homepage surfaces. Since the baseline systems already employ Semantic ID token-based embeddings, these reported metrics isolate the pure performance lift generated solely by adding the SiDec component. Furthermore, at YouTube's massive scale, these absolute percentage improvements in satisfied engagement are considered highly statistically significant, representing substantial shifts in daily user behavior. Empirical evaluations demonstrated that these architectural upgrades disproportionately benefit nascent accounts with sparse interaction histories and long-tail content, successfully alleviating traditional popularity bias.

\subsection{Breaking the I/O Bottleneck: Quality vs. Efficiency}

To evaluate the exact trade-offs between representation fidelity, retrieval capability, and computational throughput, we designed a comprehensive offline study utilizing the retrieval model. The study aims to quantify the trade-offs between system I/O efficiency (training speed) and information loss (predictive quality) when shifting from dense embeddings to SiDec token sequences.

We established five experimental arms to isolate the impact of direct embedding ingestion versus on-the-fly codebook decoding:
  
  \begin{itemize}
      \item \textbf{Control:} The baseline production model utilizing standard IDs and lightweight context features, with no pre-trained content embeddings or codebooks.
      \item \textbf{Arm 1 (raw 64-dim content embedding):} The model directly ingests a 64-dimensional pre-trained content embedding for each video from the training data. This arm represents the upper bound of I/O pressure.
      \item \textbf{Arm 2 (SID v0):} Utilizes codebook v0 to decode stored Semantic IDs into 64-dimensional latent content embeddings on-the-fly.
      \item \textbf{Arm 3 (SID v1):} Utilizes codebook v1 to decode Semantic IDs into a richer, higher-fidelity 256-dimensional latent space.
      \item \textbf{Arm 4 (SID v1 + Scaling):} Utilizes codebook v1 (256-dim) coupled with architectural scaling in the Transformer block.
  \end{itemize}

\begin{table*}[htbp]
  \caption{Performance and I/O Efficiency Comparison Across Experimental Arms with Retrieval Model.}
  \label{tab:io_study}
  \centering
  \renewcommand{\arraystretch}{1.2}
  \begin{tabular}{@{}lcccc@{}}
      \toprule
      \textbf{Experiment Arm} & \textbf{Content Representation} & \textbf{Loss @ Convergence} & \textbf{Hit Rate @100} & \textbf{Training Speed (steps/s)} \\
      \midrule
      \textbf{Control} & - & 2.766 & 0.2811 & 16.80 \\
      \textbf{Arm 1 (Raw 64-dim Embedding)} & Direct Dense Embeddings & 2.723 & 0.2844 & 12.07 \\
      \textbf{Arm 2 (SID v0)} & Codebook v0 Decoder & 2.764 & 0.2816 & 15.41 \\
      \textbf{Arm 3 (SID v1)} & Codebook v1 Decoder & 2.758 & 0.2870 & 15.26 \\
      \textbf{Arm 4 (SID v1 + Scaling)} & Codebook v1 + Scaling & 2.681 & 0.2910 & 14.53 \\
      \bottomrule
  \end{tabular}
\end{table*}

\paragraph{\underline{Baseline Analysis (Control vs. Arm 1)}} 
The evaluation (Table~\ref{tab:io_study}) confirms a fundamental structural tension between downstream quality and I/O efficiency. The Control arm, operating completely without content representation, establishes the throughput upper bound at 16.80 steps/s, but suffers from severe underfitting. In contrast, introducing raw content embeddings directly (Arm 1) infuses the model with explicit dense semantic signals, lifting the Hit Rate @100 to 0.2844. However, continuous ingestion of uncompressed dense arrays triggers a critical I/O bottleneck, degrading training throughput by 28.2\% down to 12.07 steps/s.

\paragraph{\underline{Using Codebooks (Arm 2 vs. Arm 3)}}
Discrete quantization successfully breaks this bottleneck. Arm 2 recovers throughput up to 15.41 steps/s (+27.7\% over Arm 1) while maintaining a competitive convergence loss relative to the Control. Expanding the codebook resolution in Arm 3 (SID v1) reclaims lost representation capacity entirely, elevating the Hit Rate @100 to 0.2870 while safely maintaining an agile training speed of 15.26 steps/s.

\paragraph{\underline{Synergistic Scaling Effects (Arm 4)}}
By pairing the discrete v1 codebook layer with expanded capacity inside the Transformer blocks, Arm 4 achieves the lowest global loss (2.681) and the highest candidate retrieval quality (Hit Rate @100 of 0.2910) among all testing configurations. Crucially, despite the added architectural parameters, Arm 4 completes updates at 14.53 steps/s—a 20.4\% throughput acceleration over the continuous embedding approach (Arm 1). This confirms that discrete tokenization transforms an unmanageable I/O bottleneck into an efficient compute footprint, allowing for concurrent scaling of both model depth and retrieval accuracy.

\subsection{Ablation Studies and Feature Analysis}

To isolate the specific components driving these performance gains, we conducted ablation studies on the Ranking model. The baseline for these comparisons utilized the dual-stream framework for all Semantic IDs.

First, we evaluated limiting the content reconstruction stream across different types of Semantic IDs, as well as the impact of the lightweight semantic decoder $f_\theta$ defined in Equation~\ref{eq:sidec}.

\begin{table}[h]
  \caption{Ablation studies of SiDec on different Semantic IDs in Ranking Model. Change in CTR AUC is the absolute percentage point.}
  \label{tab:ranker_1}
  \centering
  \begin{tabular}{lc}
    \toprule
    \textbf{Semantic Decoder Type} & \textbf{CTR AUC}\\
    \midrule
    Candidate \& Watch SIDs Only  & -0.04\% \\
    Watch History SIDs Only & -0.03\% \\
    No lightweight decoder & -0.01\% \\
  \bottomrule
\end{tabular}
\end{table}

Table~\ref{tab:ranker_1} demonstrates that the majority of improvements stem from the inclusion of watch history Semantic IDs, which aligns logically with the fact that watch history contains a higher volume of videos, fully leveraging the I/O efficiency of SiDec compared to single candidate videos. Removing the lightweight decoder $f_\theta$ only decreases performance slightly, suggesting that the raw decoded embeddings in the latent space provide a robust baseline signal independently.

Next, we assessed the reliance of the reconstruction stream on surrounding model architectures and feature interactions.

\begin{table}[h]
  \caption{Ablation studies on content reconstruction stream in Ranking Model. Change in CTR AUC is the absolute percentage point.}
  \label{tab:ranker_2}
  \centering
  \begin{tabular}{lc}
    \toprule
    \textbf{Ablation Type} & \textbf{CTR AUC}\\
    \midrule
    Ablate SSL  & -0.05\% \\
    Ablate Cross Attention  & -0.08\% \\
    Ablate multi-modal content & -0.06\% \\
  \bottomrule
\end{tabular}
\end{table}

The results presented in Table~\ref{tab:ranker_2} demonstrate that while the SiDec approach serves as a highly efficient representation, it requires deep integration with advanced content learning architectures—such as contrastive loss (SSL) \cite{oord2018representation} between watch and candidate videos, cross-attention mechanisms \cite{rashed2022context}, and multi-modal content representations \cite{he2026plum}—to achieve its full predictive potential.

\section{Conclusions and Future Work}

In this paper, we introduced the Dual-purpose Semantic ID framework, a novel paradigm designed to break through the "Memory Wall" and severe high-bandwidth I/O bottlenecks that restrict the scalability of modern deep recommendation systems. By leveraging hierarchical vector quantization, we compress high-dimensional, continuous content embeddings into highly compact, discrete token sequences. This design successfully shifts the system-level burden from costly, disk-bound dense vector retrieval to highly efficient, compute-bound on-the-fly reconstruction within accelerator memory.

Our dual-purpose approach extracts maximum utility from these discrete tokens by serving two concurrent, vital functions directly inside the recommendation model:
\begin{enumerate}
    \item \textbf{Collaborative Identity:} Semantic IDs act as structured categorical features mapped to hierarchical, learnable token-level embedding tables, enabling powerful collaborative filtering capabilities that naturally generalize to cold-start and long-tail items.
    \item \textbf{Content Reconstruction:} A lightweight, task-aligned Semantic Decoder reconstructs high-fidelity approximations of the original dense content embeddings on-the-fly, entirely bypassing the need to store, log, or join wide floating-point vectors.
\end{enumerate}
Our real-world deployment in production-scale ranking and retrieval systems at a major video sharing platform has demonstrated the immense practical value of this framework. We observed significant improvements in online satisfied user engagement, particularly when utilizing ultra-long user watch histories where traditional dense features would otherwise trigger prohibitive memory and serving latencies.

The philosophical takeaway of this work extends far beyond replacing traditional catalog IDs or content embeddings. Our results suggest a broader, highly promising system paradigm: \textbf{tokens are indeed all you need}. 
Almost any high-dimensional, dense continuous feature in recommendations—including complex user contextual states, spatial representations, historical engagement densities, and multi-modal feature vectors—can be quantized into a discrete token space. 
By representing the entire feature space under a unified vocabulary of discrete tokens, recommendation systems can completely decouple from continuous floating-point I/O, optimizing for pure symbolic routing and on-demand reconstruction. This aligns recommendation systems with the highly optimized, compute-bound hardware scaling laws enjoyed by Large Language Models.

Moving forward, we plan to explore several key research directions to push the boundaries of this paradigm:
\begin{itemize}
    \item \textbf{Advanced Ultra-Compression Architectures:} We intend to investigate alternative quantization strategies, such as adaptive codebook learning and multi-stage neural compression models, aiming for even higher compression ratios (e.g., $>100\times$) with near-zero downstream information loss.
    \item \textbf{Dynamic Codebook Adaptation:} To address collaborative and semantic drift, we plan to design or leverage mechanisms that dynamically update codebook boundaries without requiring a full retraining of the downstream model.
\end{itemize}

\begin{acks}
The authors would like to thank our colleague Sophia Andrikopoulos for her collaboration and contribution on the Nested N-Gram approach. We also appreciate the critical suggestions and technical support from our colleagues Lukasz Heldt, Vince Gatto and Nikhil Mehta.
\end{acks}

\bibliographystyle{ACM-Reference-Format}

\bibliography{main}

\end{document}